%% file: scubasf_astroph.tex
\documentclass{aa}
\usepackage{graphicx}
\usepackage{txfonts}
\usepackage{rotating}
\usepackage{natbib}
\usepackage{colortbl}
\usepackage{longtable}
\usepackage{lscape} % or pdflscape for pdf
\bibpunct{(}{)}{;}{a}{}{,}
\newcommand{\micron}{\hbox{$\mu{\rm m}$}}

\newcommand{\msun}{\mbox{$M_{\odot}$}}
\newcommand{\prob}{\hbox{P}}

\begin{document}
\title{Star formation in Perseus}
\subtitle{Clusters, filaments and the conditions for star formation}
\author{J. Hatchell\inst{1,2}, J. S. Richer\inst{3}, G. A.
Fuller\inst{4}, C. J. Qualtrough\inst{3}, E. F. Ladd\inst{5}, C. J.
Chandler\inst{6}}
\authorrunning{Hatchell et al.}
\titlerunning{SCUBA Perseus survey}
   
\offprints{hatchell@astro.ex.ac.uk}

\institute{Max-Planck Institut f\"ur Radioastronomie, Auf dem
   H\"ugel 69, 53121 Bonn, Germany  
\and School of Physics, University of Exeter, Stocker Road, Exeter
   EX4 4QL, U.K.
\and Cavendish Laboratory,
Madingley Road,
   Cambridge CB3 0HE, U.K.
\and Department of Physics, UMIST, P.O. Box 88, Manchester M60 1QD, U.K.
\and Department of Physics, Bucknell University, Lewisburg, PA 17837, U.S.A.
\and National Radio Astronomy Observatory, P.O. Box O, Socorro, NM
   87801, U.S.A.}

\date{}
 
\abstract{We present a complete survey of current star formation in
   the Perseus molecular cloud, made at 850 and 450~\micron\ with
   SCUBA at the JCMT.  Covering 3~deg$^2$, this submillimetre continuum
   survey for protostellar activity is second in size only to that of
   $\rho$~Ophiuchus \citep{johnstone04}.  Complete above 0.4~\msun\
   ($5\sigma$ detection in a $14''$ beam), we detect a total of 91
   protostars and prestellar cores.  Of these, 80\% lie in clusters,
   representative of star formation across the Galaxy.  Two of the groups
   of cores are associated with the young stellar clusters IC348 and
   NGC1333, and are consistent with a steady or reduced star formation
   rate in the last 0.5~Myr, but not an increasing one.  In Perseus,
   40--60\% of cores are in small clusters ($< 50$~\msun) and isolated
   objects, much more than the 10\% suggested from infrared studies.
   Complementing the dust continuum, we present a C$^{18}$O map of the
   whole cloud at $1'$ resolution.  The gas and dust show filamentary
   structure of the dense gas on large and small scales, with the
   high column density filaments breaking up into clusters of cores.
   The filament mass per unit length is 5--11~\msun\ per 0.1~pc.
   Given these filament masses, there is no requirement for substantial
   large scale flows along or onto the filaments in order to gather
   sufficient material for star formation.  We find that the probability
   of finding a submillimetre core is a strongly increasing function
   of column density, as measured by C$^{18}$O integrated intensity,
   $\prob(\hbox{core}) \propto I\,^{3.0}$.  This power law relation holds
   down to low column density, suggesting that there is no $A_{\rm v}$
   threshold for star formation in Perseus, unless all the low-$A_{\rm
   v}$ submm cores can be demonstrated to be older protostars which have
   begun to lose their natal molecular cloud.

}

\maketitle

\section{Introduction}

Recent advances in submillimetre (submm) detector technology now make
it possible to image entire star forming regions and detect all the
protostellar and starless cores within.  For the first time, we can
gather statistically significant samples of stars at the earliest
stages of their evolution.  At the same time, theories of star
formation are moving on from the formation of individual, isolated
objects \citep[e.g.,][]{shuadamslizano87} to the formation of clusters
\citep[e.g.,][]{maclow04rev}.  It is thus hugely important at this time to
make large surveys of star formation regions to compare with the theories.

An important question which we can only address with submm surveys
is where in molecular clouds stars form.  This cannot be answered by
looking at older populations of pre-main-sequence stars detectable in
the infrared and optical because the typical velocities of the stars
lead them to wander by 1~pc/Myr, and the environment in which they formed
   may no longer exist.
Therefore we are driven to look
   for
stars at the point of formation or shortly after, at protostellar cores
and deeply embedded Class~0 objects, with ages less than or of order
$10^4$ years.  With submm surveys we can relate the positions of the
young protostars to the properties of the dust and gas in the molecular
clouds in which they form.  This has led other authors \citep{onishi98,
johnstone04} to conclude that there is an extinction or column density
threshold for star formation, that is, that stars only form in the
densest parts of molecular clouds.
   We investigate this conclusion further in this paper.

We have carried out a complete survey of star formation above 0.4~\msun\
in the Perseus molecular cloud.  Perseus was selected as it is known to
be forming clusters of stars and more massive stars than, e.g., Taurus.
The Perseus molecular cloud contains a total of $17,000$~\msun\ of gas,
estimated from
   visual
extinction \citep{bachiller86av}.  Our survey area of $\sim 3$~deg$^2$
was selected to include all the moderate to high extinction material
($A_{\rm v} > 4$) where star formation might occur.  Imaging dust
emission at 850 and 450~\micron\ with the Submillimetre Common User
Bolometer Array (SCUBA) at the James Clerk Maxwell Telescope (JCMT), we
detect all the protostars and starless cores above 0.4~\msun/14$''$~beam.
Complementing the dust continuum images, we
   present
molecular line (C$^{18}$O~$J$=1--0) data at $1'$ resolution.  Thus,
we can relate the star formation activity to the underlying physical
properties of the molecular cloud and to the kinematics.  By imaging a
single molecular cloud we minimise the uncertainties in derived masses,
sizes, etc., due to distance.

The Perseus molecular cloud is a well known star forming cloud in
the nearby Galaxy.   It is associated with two clusters containing
pre-main-sequence stars: IC348, with
   %%
   %% a Hipparcos distance
   %% of 320~pc and
   %%
an estimated age of 2~Myr (and spread of $\pm 1.5$~Myr;
\citealp{luhman03}); NGC1333,
   %%
   %% at 318~pc
   %%
which is less than $1$~Myr in age \citep{lal96,wilking04}; and the
Per~0B2 association, which contains a B0.5 star \citep{steenbrugge03}
and therefore must be less than 13~Myr in age \citep{meynetmaeder00}.
These clusters therefore all show evidence of
   %%
   %% earlier
   %%
star formation activity within the last $\sim 10^7$ years.  The molecular
cloud itself contains a number of previously known protoclusters and
isolated protostars but until now there has been no complete census of
the current star formation activity.

   We assume the distance to the Perseus molecular cloud to be the same
   as the Hipparcos distance to the embedded clusters, and use 320~pc
   throughout \citep{dezeeuw99}.

\section{Observations and data reduction}
\label{sect:observations}

\begin{figure*}[p]
\centering
\includegraphics[scale=0.70,angle=-90]{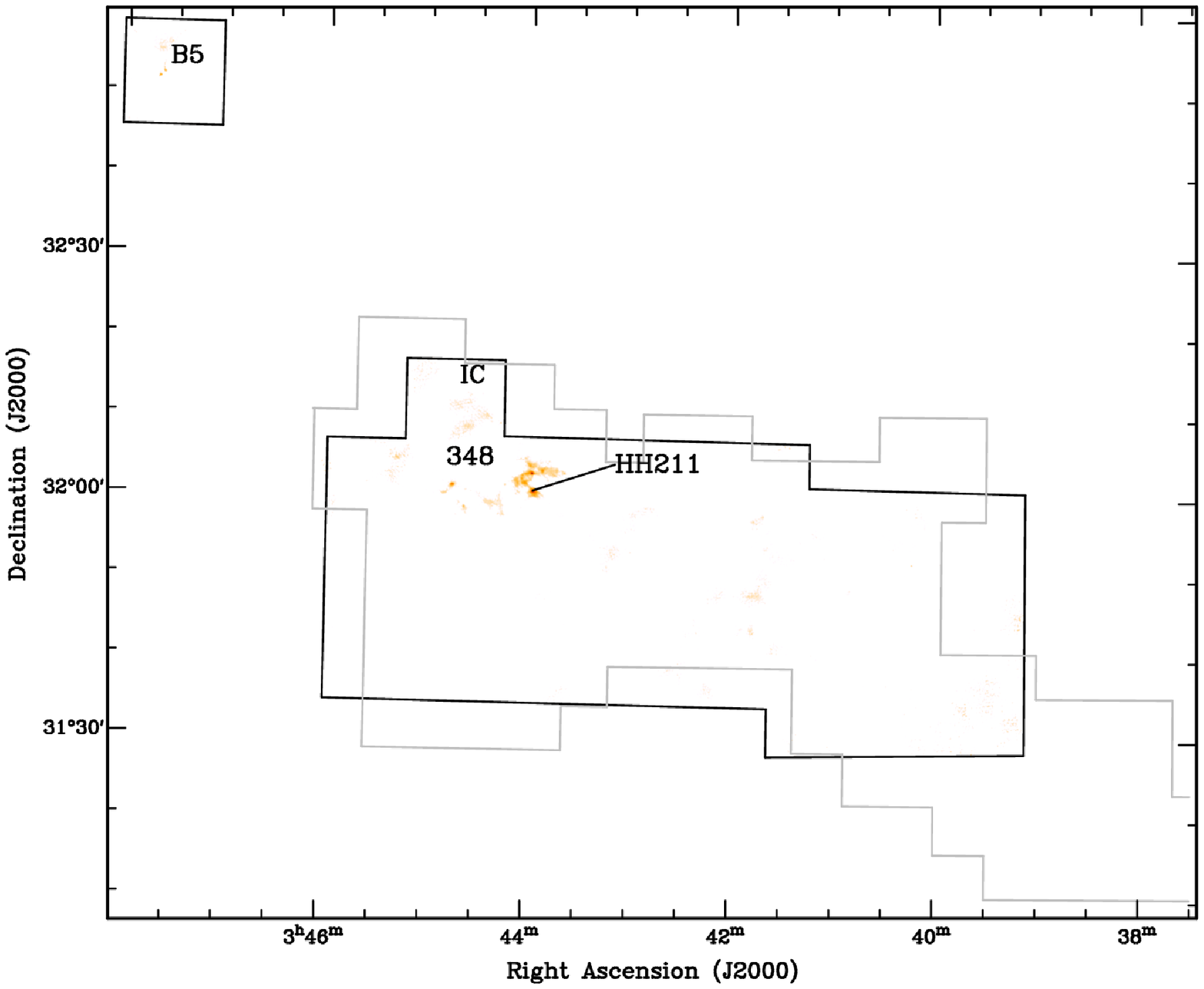}\\
\includegraphics[scale=0.70,angle=-90]{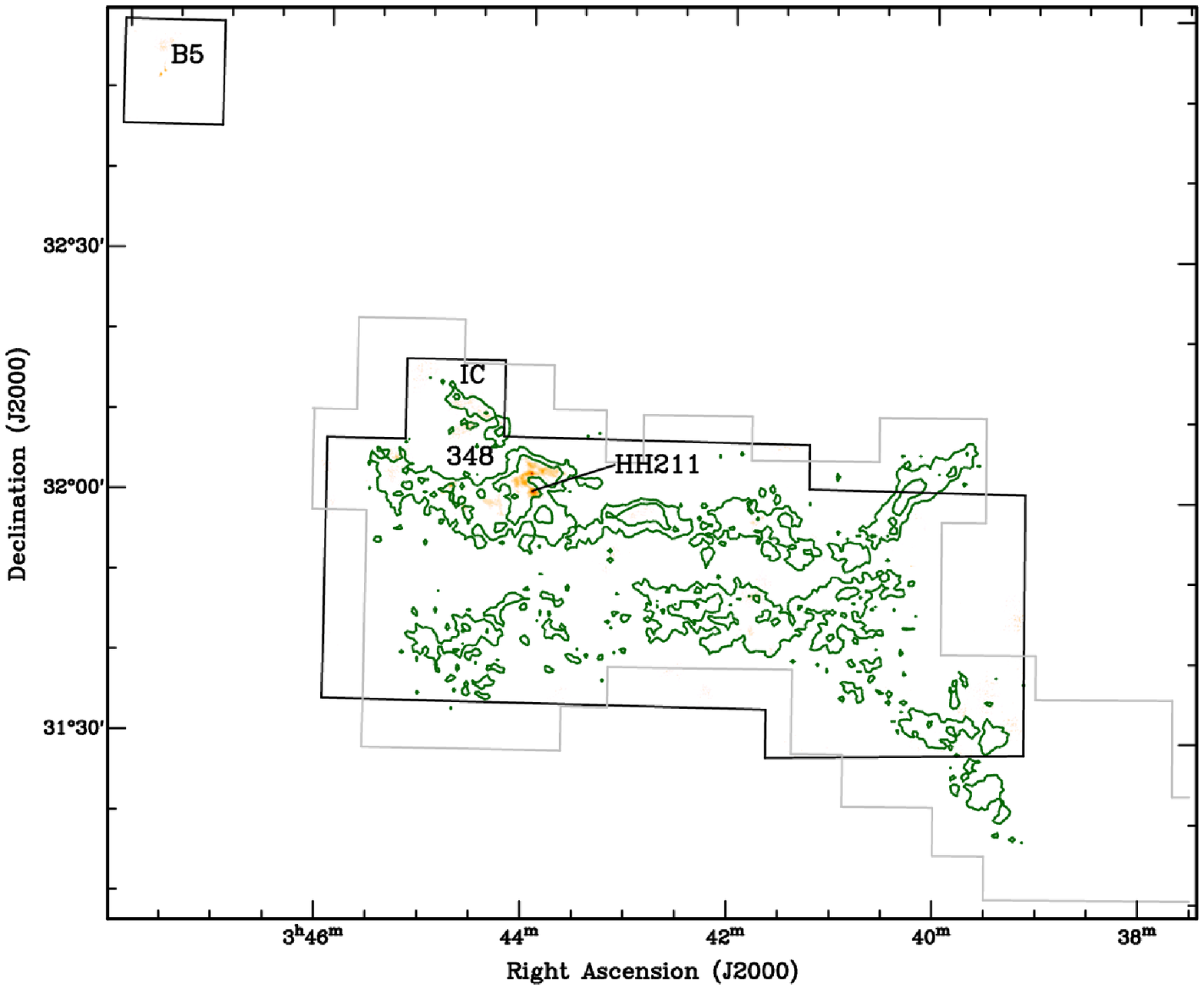}\\
\caption{{\bf a.} The eastern part of the Perseus molecular cloud
   complex.  {\bf Top:} SCUBA 850~\micron\ map.  The colourscale is from
   0 to 2281~mJy/beam on a square root (flux density) scale to bring out
   the low intensity features.  {\bf Bottom:} Contours of C$^{18}$O~1--0
   integrated intensity at $1'$ resolution overlaid on 850~\micron\
   colourscale.  Contours are 1, 2, 4, 8, 16~K~km~s$^{-1}$.  In both maps
   the black boundary marks the area mapped with SCUBA and the grey
   boundary marks the area mapped in C$^{18}$O (note, B5 was not mapped
   in C$^{18}$O).  Some well-known clusters and IRAS sources are marked.}
\label{fig:scubaandCO}
\end{figure*}

\setcounter{figure}{0}

\begin{figure*}[p]
\centering
\includegraphics[scale=0.84,angle=-90]{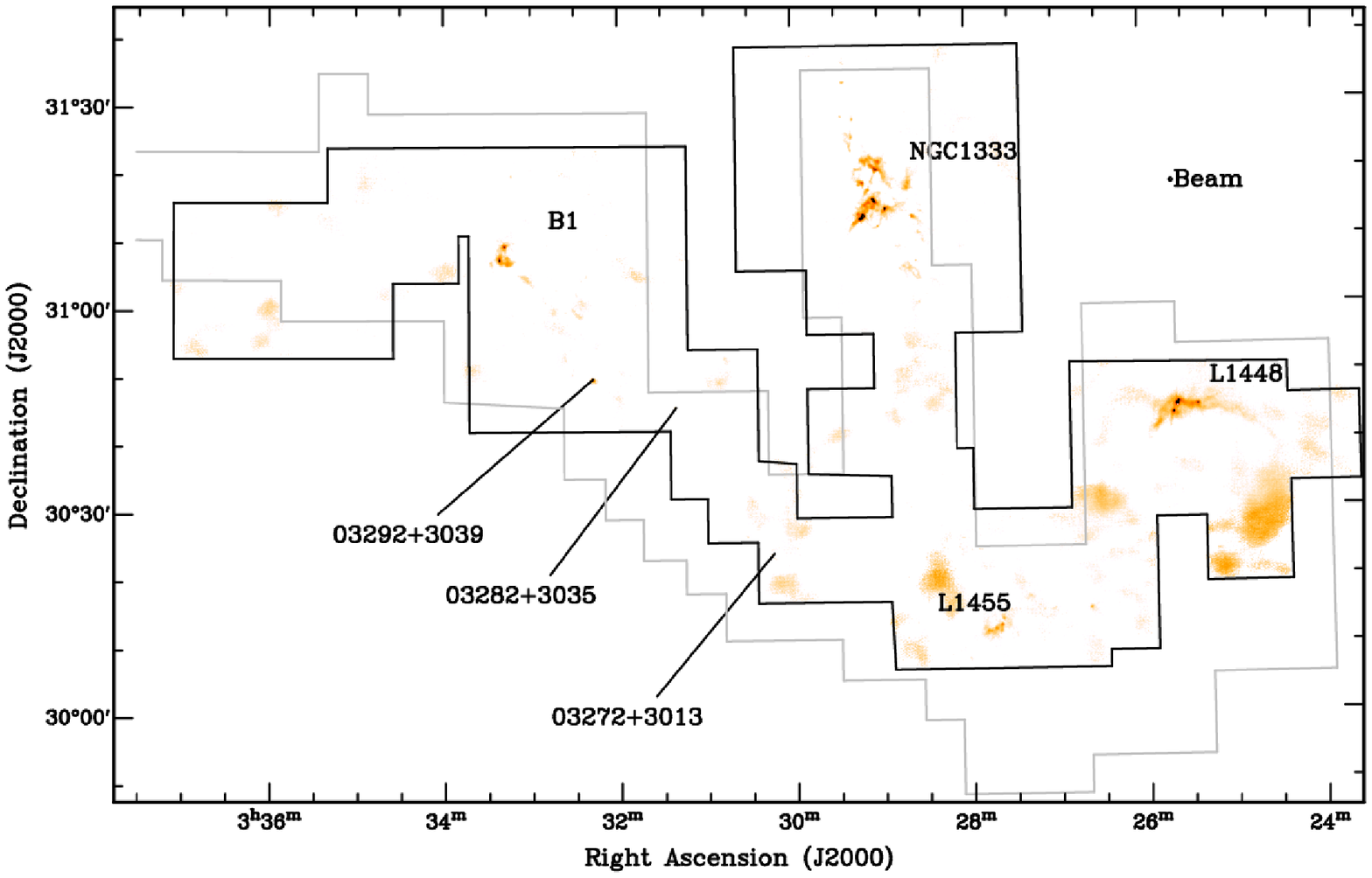}\\
\includegraphics[scale=0.84,angle=-90]{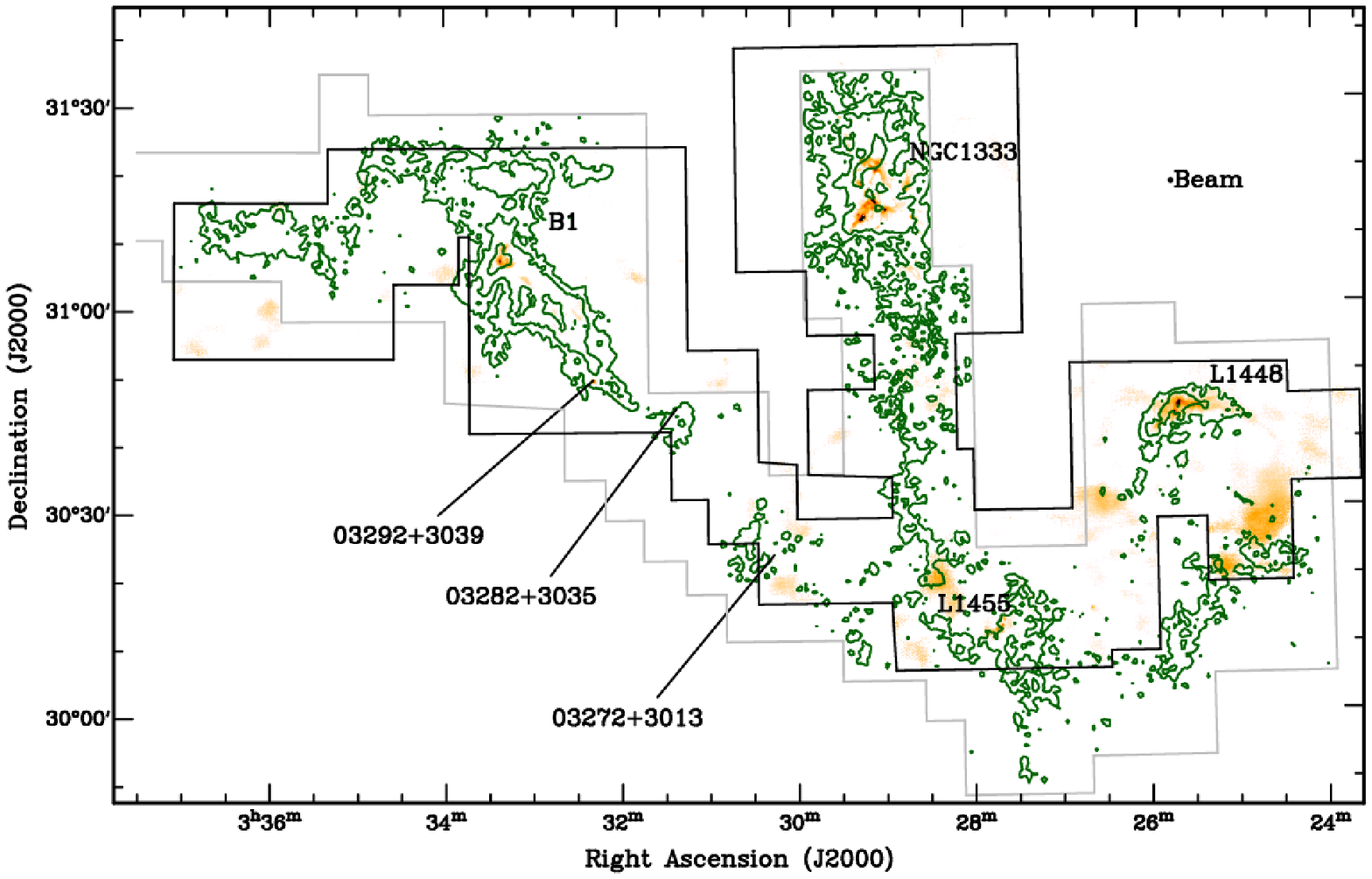}\\
\caption{{\bf b.} The western part of the Perseus molecular cloud
   complex.  {\bf Top:} SCUBA 850~\micron\ map.  The colourscale is from
   0 to 2281~mJy/beam on a square root (flux density) scale to bring out
   the low intensity features.  {\bf Bottom:} Contours of C$^{18}$O~1--0
   integrated intensity at $1'$ resolution overlaid on 850~\micron\
   colourscale.  Contours are 1, 2, 4, 8, 16~K~km~s$^{-1}$.  In both maps
   the black boundary marks the area mapped with SCUBA and the grey
   boundary marks the area mapped in C$^{18}$O (note, B5 was not mapped
   in C$^{18}$O).  Some well-known clusters and IRAS sources are marked.}
\end{figure*}

\begin{sidewaysfigure*}[p]
\centering
\includegraphics[scale=0.85,angle=-90]{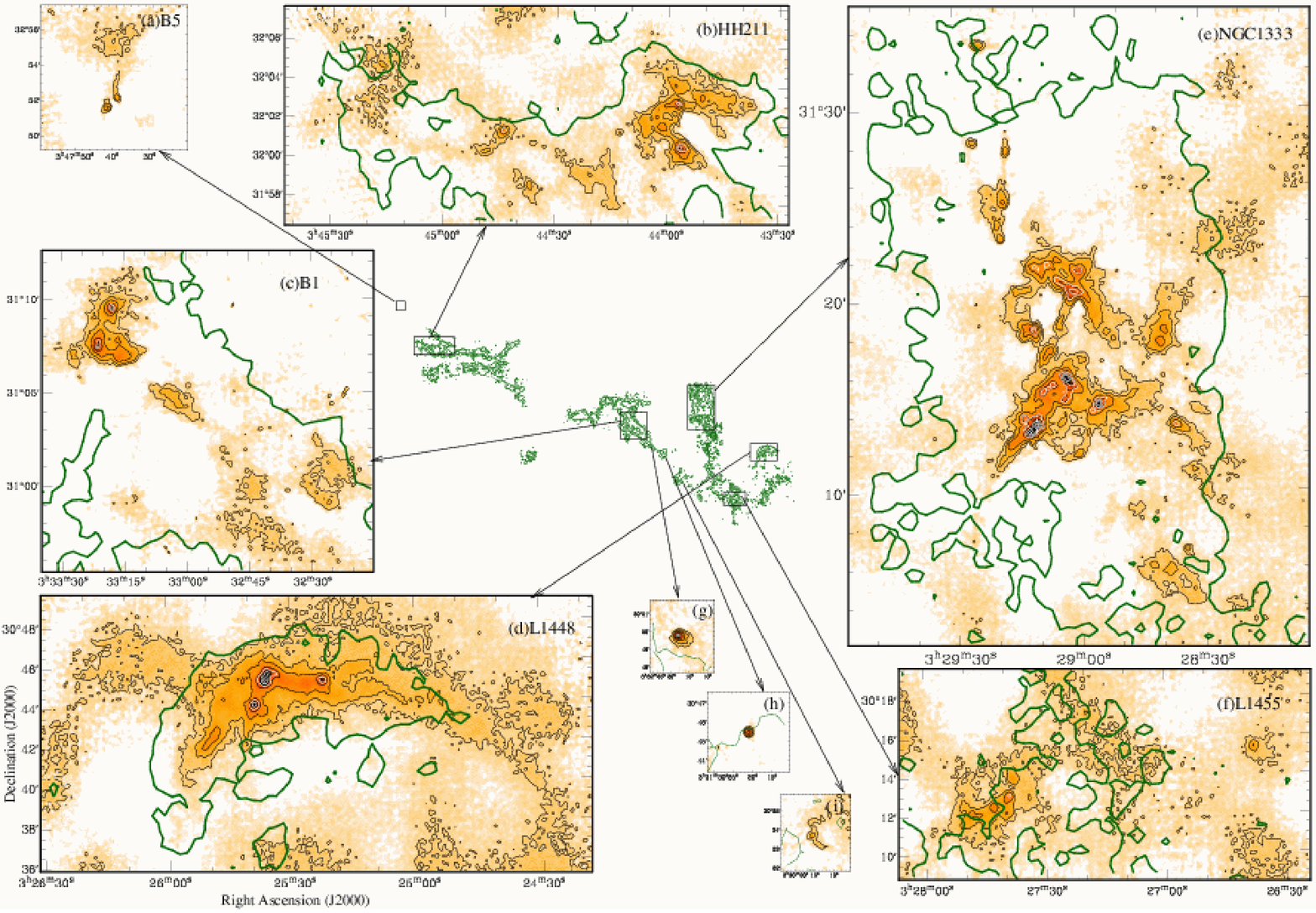}\\
\caption{SCUBA 850~\micron\ maps of the (a) B5, (b) HH211, (c) B1, (d)
   L1448, (e) NGC1333, (f) L1455, (g) IRAS~03292+3039, (h) IRAS~03282+3035
   and (i) IRAS~03271+3213 regions.  Colourscale is 850~\micron\ continuum
   from 0 (white) to 2281~mJy/beam (black) on a square root (flux density)
   scale to bring out the low intensity features.  Black contours are
   850~\micron\ 100, 200, 400, 800, 1600, 3200~mJy/beam.  White contours
   are 450~\micron\ 2, 4, 8~Jy/beam.  Dark green contours are where the
   C$^{18}$O integrated intensity = 1~K~km~s$^{-1}$.}
\label{fig:clusters}
\end{sidewaysfigure*}

Images of Perseus were obtained at 850 and 450~\micron\ using SCUBA on
the JCMT
   during
20 nights between 1999 and 2003.  The region selected
corresponds roughly to the $\int T_{\rm A}^*\,dV =
4\hbox{ K km s}^{-1}$ contour of the Bell labs 7m $^{13}$CO
map\footnote{The $^{13}$CO map is available from John Bally, at \\
\texttt{http://casa.Colorado.EDU/$\sim$bally/ftp/7mdata/\\Perseus\_13co.fits}},
although not all of this area was mapped with both SCUBA and C$^{18}$O,
while some regions outside this contour were mapped with each.
For completeness and calibration consistency we
   included
the NGC1333 region although it was mapped previously with SCUBA by
\citet{sandell&knee01}.  The whole region
   covered
contains more than 10,000~\msun\ of gas, estimated from $^{13}$CO.
Fields of size $10'\times 10'$ were scan-mapped 6 times with each of 3
chop throws ($30''$, $44''$ and $68''$) and 2 chop directions (R.A. and
Dec.).  The data were reduced using SURF \citep{surf}.  The typical RMS
noise level is 35~mJy/beam at 850~\micron.  The
   weather
conditions for
   the
450~\micron\ observations were more variable: for the clusters
(observed in the best conditions as we expected to find sources here)
the noise level is $\sim 200$~mJy/beam, but increases to $> 1$~Jy/beam
in other regions.  The 450 and 850~\micron\ beam sizes are $8''$ and
$14''$ respectively or 0.012 and 0.022~pc at the distance of Perseus.
We assume a dust emissivity of $\kappa_{850} = 0.012$~cm$^2$~g$^{-1}$
(gas+dust), from \cite{oh94} model 5 for icy coagulated grains and
a gas-to-dust mass ratio of 161 ($1.5\times 10^{-26}$~g of dust per
H atom; \citealp{draine&lee84}), assuming gas is 0.89\% H by number
\citep{cww90}.  Our beam mass sensitivity to a 12~K core is then
0.4~\msun\ at 850~\micron, although we also detect warmer, less massive
objects, and our beam-averaged column density sensitivity is $6\times
10^{22}$~cm$^{-2}$ (both limits given for $5\sigma$ detections).

Maps of the molecular cloud in C$^{18}$O~$J$=1--0 were made in January
and December 2000 using the $4\times4$ array SEQUOIA on the FCRAO 14m
telescope.  The beamsize was $46''$, corresponding to 0.07~pc at the
distance of Perseus.  The 92,500 spectra were baselined, calibrated to
$T_{\rm MB}$ with an efficiency of 0.45, smoothed to 0.25~km~s$^{-1}$
resolution, gridded and integrated
   in
velocity over 0--12~km~s$^{-1}$ using the GILDAS software, to produce
the integrated intensity map shown in Fig.~\ref{fig:scubaandCO}.
Visualisation and further analysis of the C$^{18}$O were carried out
using the Karma visualisation suite \citep{gooch96}.
   Maps of $^{13}$CO(1--0) were also obtained during this same observing
   period, the results from which will be presented at a later date.

We take
   %%
   %% (this is now moved to the end of the intro -- CJC)
   %%
   %% 320~pc for the distance of the molecular cloud (the Hipparcos
   %% distance to the embedded clusters; \citealp{dezeeuw99}), and
   %%
$N$(C$^{18}$O$) = 3.4\times 10^{14}(A_{\rm v}-1.6)$~cm$^{-2}$
\citep{bachiller86av}, or alternatively, $I$(C$^{18}$O$) =
(A_{\rm v}-1.6)/3.6$~K~km~s$^{-1}$, assuming LTE at 12~K.  We also
assume $N$(H+2H$_2) = 0.9\times 10^{20} A_{\rm v}$ \citep{bohlin78}.
The temperature of the molecular gas is typically 10--12~K from NH$_3$
observations \citep{bachiller86nh3}, although protostellar cores are
warmer (30--50~K: \citealt{motteandre01,jennings87}).

Our definition of a core for the purposes of this paper is a flux
peak with closed contours at the 500, 250 or 150~mJy ($\sim 4\sigma$)
levels.  Elongated regions at those levels were divided into subpeaks at
subsidiary maxima.  It should be borne in mind that it is not necessarily
straightforward to convert our observational sensitivity of 0.4~\msun/beam
to the minimum total mass of a core identifiable in our survey as this
limit depends on the core density distribution and size: massive but
extended cores could easily remain undetected.  As a consequence it is
likely we are least sensitive to the youngest and hence least centrally
condensed prestellar cores.  In addition if the pressure external to a
core determines its size for a given mass, our core mass sensitivity
may vary from region to region in the cloud giving a higher limit in
the lower pressure regions away from the clusters \citep{johnstone04}.
We are also insensitive to weak sources in the emission wings of strong
ones and therefore underestimate the number of weak sources in clusters,
an effect which is discussed further in the next section.

\clearpage

\section{Clusters}
\label{sect:clusters}

The complete 850~\micron\ image in Fig.~\ref{fig:scubaandCO} shows
emission from dust cores and high column density gas.  In many cases the
cores already show central concentration indicative of gravitational
collapse, and must be associated with the earliest stages of star
formation.

We detect 91 dust cores with an 850~\micron\ flux above 150~mJy/beam
($> 4\sigma$ detections). 
%% GAF 
These are
listed in
Table~\ref{tbl:sourcelist}.  About half of these are new submm
detections.  We see from Fig.~\ref{fig:scubaandCO} that most of the
cores lie in five main groups: from east to west in the cloud, the
HH211 region, B1, NGC1333, L1455 and L1448.  Additionally, there are a
few scattered isolated cores and two cores in B5.  These groups are
shown in more detail in Fig.~\ref{fig:clusters}, where the individual
cores and the filamentary structure connecting them are more evident.
A full inventory of the pre/protostellar population of Perseus
including estimates of individual core masses will form the basis of a
later publication, but it is clear from the submm fluxes that these
groups contain up to a few solar masses of stars.

Lower flux sources appear slightly less clustered than stronger sources
(78\% of sources
   with $F_{\rm 850~\mu m} < 250$~mJy are in clusters compared to 88\%
   of sources with $F_{\rm 850~\mu m} > 500$~mJy),
but this may be due to our core selection technique, which is less
sensitive to weak cores in clusters (see Sect.~\ref{sect:observations}).
While there may be additional physical reasons why higher mass (and
therefore brighter submm) sources  preferentially form in clusters,
we are not yet in a position to test this.

Making a count of cores, we find that more than 80\% of the cores fall
into five main clusters. 
   We define a core 
to be in a cluster if it has 2 or more neighbours within 0.5~pc.
With our sensitivity to all cores above 0.4~\msun,  this definition
guarantees a mass surface density of these regions of at least 1.2~\msun\
within 1~pc$^{2}$.  This
   definition is then consistent with the Spitzer tidal stability criterion
   that a stellar density $>1.0$~\msun~pc$^{-3}$ is required in order
   for a cluster not to be disrupted on a timescale $\sim 10^8$~yr
   \citep{spitzer58}, although during star formation clusters
almost certainly rely on the gravitational potential of the molecular gas
for binding. The remaining few isolated cores are scattered throughout
the molecular cloud.  80\% is probably a lower limit on the fraction of
star formation in clusters in Perseus because we underestimate the number
of cores in clusters due to confusion (see below).  Infrared studies of
embedded clusters of pre-main-sequence stars estimate the fraction of
stars in clusters as 60--90\% (L1630; \citealp{ealada91}) and 50-100\%
(4 nearby clouds; \citealp{carpenter2000}),  which are consistent with
our findings.

This clustered mode of star formation contrasts with the Taurus region
\citep{andremontmerle94}, where isolated star formation dominates.
With the majority of stars in the Galaxy forming in clusters
\citep{ladalada03}, Perseus, like Orion~B \citep{mitchell01} is more
%% GAF
likely
representative of typical star formation in our Galaxy.

\subsection{Core clusters and stellar clusters}

Two of the groups of cores are spatially associated with young stellar
clusters: the HH211 group can be associated with IC348, and the NGC1333
cores with the NGC1333 cluster, and may be evidence for continuing star
formation in these clusters.
   We now
consider the evolution of star formation in these regions by comparing
the number of stars in the clusters with the likely stellar yield of
the current protostellar cores.

As our pre/protostellar core sample is incomplete below 0.4~\msun, we take
this as the lower mass limit for the count.  The yield (in stars above
0.4~\msun) of the protostellar cores can only be a rough estimate because
of the unknown multiple star fraction and the unknown final masses of the
accreting stars.  We assume each core yields between 1 and 2 stars above
0.4~\msun; this takes into account complex cores forming multiple stars.

 In IC348, the infrared population is consistent with continuous
   star formation at a rate of 50 stars
with masses
above 0.4~\msun\ per Myr from 3.5~Myr to 0.5~Myr ago \citep[based on
an estimate that 2/5 of the total of 348 stars shown on the IMF in
their Fig.~16, have masses above 0.4~\msun]{muench03}, This value and
a constant star formation rate, suggests $\sim 25$ embedded protostars
would be expected, if the embedded phase has a lifetime of 0.5~Myr.
This is consistent with the 18 submm cores, containing an estimated
18--35 protostars, we detect in the HH211 region.

   Taking an age spread and population for NGC1333 of 0.5--1.5~Myr
   and 143 pre-main-sequence stars \citep{wilking04}, a constant star
   formation rate would predict $\sim 70$ protostars in the last 0.5~Myr,
   whereas we observe 36 cores containing an estimated 36--70 protostars.
   Here the submm population appears consistent or
a little
   less than that expected from a constant star formation rate and an
   embedded phase lifetime of 0.5~Myr.
   
   An embedded phase lifetime of 0.5~Myr is consistent with the
   youngest T~Tauri sources in IC348 and NGC1333, but is longer than
   estimates for the embedded phase lifetime in $\rho$~Ophiuchus and
   Taurus-Auriga, 0.1--0.4~Myr \citep{wilking89,kenyon90}.
%%
%%    The conclusions would change little
%%    if we took an embedded phase lifetime at the higher end of this
%%    range, 0.3--0.4~Msun.  In NGC1333, this shorter timescale for the
%%    embedded phase would better match the observed T~Tauri/embedded
%%    star ratio. 
%%
   The age spread of T~Tauri populations is notoriously difficult to
   estimate from isochrones \citep{hartmann01}, and the youngest IC348
   and NGC1333 infrared members may well be less than 0.5~Myr old.
However, an
   embedded phase lifetime as short as 0.1~Myr would require either a
   significant correction to the T~Tauri ages or that we happen to be
   looking at a star formation burst after a significant gap in both
   IC348 and NGC1333, which seems unlikely.

   In conclusion, in IC348 and NGC1333 our observations are consistent
   with a steady or reduced star formation rate over the last 0.5~Myr,
   but not an increasing one.

The small population
   of cores in 
many of the clusters in Perseus raises the question of the cluster IMF
-- how many stars form in large/massive clusters compared to small ones?
The massive NGC1333 cluster, which has a stellar population of 143 stars
with a total stellar mass of 79\msun\ \citep{lal96} contains 40\% of our
pre/protostellar cores.  A further 15\% lie in the HH211 region bordering
IC348, which has $>300$ stellar members, and 160~\msun \citep{luhman03}.
So, in Perseus, 40--60\% of cores are in massive clusters having $>
50$~\msun. 
%% GAF
On the other hand
small clusters and isolated objects also contribute
40--60\%, much more than the 10\% suggested from the infrared studies:
\citet{ladalada03} suggest that 20--50~\msun\ clusters contribute
no more than 10\% of all stars.  Unlike our statistics on cores, our
statistics on clusters are derived from a very small sample of 5, so are
less certain.  
%%GAF 
%%However, there are biases in each wavelength study which
%%should be mentioned.  
%%
%%The infrared studies used to derive these statistics
%%are typically sensitive to stars in the mass range which we would expect
%%our cores to produce ($> 0.1$--0.3\msun; \citealt{carpenter2000}), so one
%%would expect consistency between the infrared and submillimetre results.
%% GAF
Nevertheless, although the infrared and submillimetre observations are
sensitive to similar ranges of masses of stars ($> 0.1$--0.3\msun;
\citealt{carpenter2000}), the difference in age  between the
submillimetre sources and the infrared visible stars may explain this 
discrepancy.
It is quite likely that the small protostellar clusters are not stable and
therefore do not classify as clusters by the Lada \& Lada criteria ($>35$
members obeying the Spitzer tidal stability criterion) by the time they
have become visible at near infrared wavelengths.  At a typical velocity
of 1~km~s$^{-1}$ a star moves 1~pc in 1~Myr or $10'$ at the distance
of Perseus.  A small group of $<10$ prestellar cores which evolved to
form a 1~Myr pre-main-sequence population scattered over a volume $20'$
in diameter could well remain unidentified as a cluster in the infrared.
Therefore infrared estimates of the amount of star formation in small
groups may be biassed
   low.
   On the other hand, in the submm, by taking a snapshot at the
   earliest stages of star formation, we underestimate the count of
   objects in clusters because future star formation is not counted.
   The smaller clusters may yet go on to form many more stars.
   Specifically, B1 has a large reservoir of molecular gas and no
   known population later than Class~I, and so may be at the earliest
   stages of cluster formation.

   The total mass of gas above a column density threshold of $6\times
   10^{22}$~cm$^{-2}$ in the clusters and cores is estimated from
   the dust emission to be $\sim 2600$~\msun.
This assumes isothermal 12~K dust: the mass will be lower if the
temperature is higher,  as is the case around the protostars.  Lowering
the flux threshold from the $5\sigma$ level to include more low column
density material would result in a higher mass.  A 5$\sigma$ cutoff is
   used to avoid contributions from artefacts on large spatial scales due
   to the reconstruction of the chopped map.
With the 5$\sigma$ cutoff, this contribution is
   limited to be
less than 30\% of the total: the mass estimate for the clusters alone
(NGC1333, HH211, B1, L1448, L1455) is 1800~~\msun.  The mass fraction of
the molecular cloud actively involved in star formation at this time is
less than 20\% of the total gas mass of 17,000~\msun\ estimated from the
   visual
extinction \citep{bachiller86av}.  The mass of dense molecular gas traced
by C$^{18}$O~1--0, which has a column density threshold for detection
of $1.6\times 10^{21}$~cm$^{-2}$ $(A_{\rm v} \simeq 2)$ and a critical
density of 400~cm$^{-3}$, is $\sim 6000$~\msun, lower than the mass
estimate from extinction as it excludes low density and atomic gas.
Thus a greater fraction of high column density gas  than low
column density gas  is in the active star formation regions,
up to as much as half of the gas above $A_{\rm v}\simeq2$.

\section{Filaments}
\label{sect:filaments}

The SCUBA 850~\micron\ image shows filamentary structure throughout the
cloud (Fig.~\ref{fig:clusters}).  The active protostellar clusters are
linked to dense filaments: the horseshoes of HH211 and B1, the arcs of
L1455 and L1448, and the zigzag of dense gas in NGC1333.  Weaker filaments
imaged by SCUBA to the east of HH211, the south of B1, the south of
NGC1333 and around L1448 each reach
   up to
1~pc in length, with the upper limit here presumably set by our
sensitivity limit.   The dust filaments have typical full width half
maxima of $1'$.  Filaments to the east of the HH211 cluster and to the
southwest of B1 do not appear to be connected to current protostellar
activity but may be fragmenting into cores to form the next round of
stars.  All the filaments show intensity fluctuations along their length,
with the most fragmented filaments already identified as strings of cores.

On much larger scales, $\sim 20$~pc, the C$^{18}$O map
(Fig.~\ref{fig:scubaandCO}), which traces material down to lower
column densities, also shows filamentary structure.  The dust
filaments detected by SCUBA are found in regions of high C$^{18}$O
column density.  In fact the entire Perseus cloud can be described as
consisting of one filament with a continuous velocity structure along
the cloud \citep{bachiller86nh3}.  SCUBA detects 850~\micron\ emission
at the 5-$\sigma$ level at H$_2$ column densities of $\sim 6\times
10^{22}$~cm$^{-2}$ ($A_{\rm v} \simeq 60$), as the C$^{18}$O starts to
become optically thick.  This column density is rarely reached and the
SCUBA map is mostly empty.   Some extended structure is also
  lost in the reconstruction of the SCUBA maps because of the limited
  range of chop throws.

The filaments are strongly reminiscent of structures formed in turbulent
flows \citep{maclow04rev}.  What we see in the dust continuum is that
the filaments are observed down to the small scales and high densities
associated with cluster formation.  This suggests that not only do
molecular clouds on the parsec scales of the whole Perseus molecular cloud
form by turbulence, but that the role of the filaments persists down to
the cluster formation scale.  A more detailed investigation of the role
of turbulence, including kinematical information from molecular lines,
is left to a future paper.

   The mass in the filaments is substantial with a mean mass per unit
   length of between 4.7 and 11.5~\msun\ per 0.1~pc ($\sim 1'$), assuming
   the filaments are isothermal at 12~K, and taking a sample of 43 filament
   cross-sections.  The uncertainty is due to the subtraction of background
   flux, some of which is extended flux belonging the filaments and some
   of which is an artefact of the reconstruction process.  An estimate
   of the artificial background in regions of blank sky suggests that
   the true filament mass lies roughly midway between these values.

The filament mass per unit length of 5--11~\msun\ per 0.1~pc ($1'$)
is 2--4 times the traditional thermal Jeans mass for a Jeans length of
0.1~pc but this does not imply that the filaments have to be unstable.
The criteria for a filament are different for those of a uniform
medium \citep{larson85}.  Theoretical models for filaments range from a
simple non-magnetic isothermal filament \citep{ostriker64} to magnetised
filaments with helical fields which can be toroidal or poloidal or both
\citep{tilley&pudritz03,fiege&pudritz00,fiege04}.  A key prediction
from these models is the ratio of the mass per unit length $m$ to the
cylindrical virial mass per unit length $m_\mathrm{vir}$, which is
given by
$$m_\mathrm{vir} = {2\langle\sigma_\mathrm{tot}^2\rangle\over G},$$
where $\sigma_\mathrm{tot}$ is the 1-dimensional total (thermal
plus non-thermal) velocity dispersion of the average gas molecule.
For filaments with or without magnetic field,
$${m\over m_\mathrm{vir}}\biggl(1-{M\over {|W|}}\biggr) = 1 -
{P_{\mathrm S}\over\langle P \rangle},$$
where $M /{|W|}$ is the ratio of magnetic energy to gravitational
energy per unit length.  Following \citet{fiege&pudritz00}~Table~2,
we assume the external pressure to be $10^{4.5}$~K~cm$^{-3}$,
and estimate the internal pressure from
$$P/k = {\rho \sigma_\mathrm{tot}^2\over k} = 2\times 10^6 \hbox{ K
cm}^{-3}$$
using an average density in the filaments of $10^5$~cm$^{-3}$.  The exact
values of these quantities are unimportant; what is important is that
 the ratio of external to internal pressure $P_{\mathrm S}/\langle
P \rangle \simeq 0.01$, which is much less than 1, so there are three
regimes: $m / m_\mathrm{vir}\sim 1$, the filament can exist without
magnetic support, $m / m_\mathrm{vir} \gg 1$ a supporting poloidal field
is required, and $m / m_\mathrm{vir} \ll 1$, when a binding toroidal field
is required (see \citealt{fiege&pudritz00,fiege04} for details).

   From a sample of 7 C$^{18}$O~1--0 spectra towards filaments,
   we measure velocity FWHM of 0.6--1.0~km~s$^{-1}$.  The conversion
   from velocity FWHM $\Delta v_\mathrm{obs}$ measured in C$^{18}$O to
   $\sigma_\mathrm{tot}$, taking into account that the C$^{18}$O molecule
   is more massive than the predominant H$_2$, is
   $$\sigma_\mathrm{tot}^2 ={1\over 8 \ln 2}\biggl(\Delta v_\mathrm{obs}^2
   + 8\ln 2{kT\over m_\mathrm{P}}\biggl({1\over \bar m} - {1\over
   m_\mathrm{C18O}}\biggr)\biggr), $$
   where $m_P$ is the proton mass, $\bar m = 2.33$ assuming 1 He for
   every 5 H$_2$ molecules, $T=10$~K, and $m_\mathrm{C18O} = 30$.

   The measured C$^{18}$O linewidths correspond to virial masses of
   45--100~\msun~pc$^{-1}$, compared to the measured mass per unit length
   of 47--115~\msun~pc$^{-1}$.  This simple analysis suggests that the
   Perseus star forming filaments
are consistent with being non-magnetic filaments \citep{ostriker64}.
This last
   result disagrees with the \citet{fiege&pudritz00} studies of
   star-forming filaments, which typically used $^{13}$CO linewidths,
   and found $m/m_\mathrm{vir} < 1$.  From a sample of 8 $^{13}$CO~1--0
   spectra, we measure velocity FWHM of 1.4--2.0~km~s$^{-1}$, typically
   a factor of 2 greater than C$^{18}$O linewidths.  If instead these
   $^{13}$CO linewidths are representative of the velocity field within
   the filaments then ours too require additional binding by a toroidal
   field.  However, it seems more likely to us that the C$^{18}$O
   linewidths are more appropriate for the thermal support within the
   dense filaments, as C$^{18}$O profiles are less affected by optical
   depth, and that the correct solution is one where the total toroidal
   field is small.

The
   %%
   %% uncertainty in the
   %%
mass derived from the dust emission, which increases if either the dust
temperature or the 850~\micron\ emissivity is lower
   than those assumed,
both of which are possibilities, adds another uncertainty.  An examination
of the transverse column density structure and the dust polarisation
would provide a further test of the filament support but
   this is beyond the scope of the current paper.

The alternative conclusion that could be reached from the high virial
mass is that the filaments are not stable but transient, 
dispersing on the crossing time of 0.1~Myr (assuming a filament width of
0.1~pc and velocity dispersion from C$^{18}$O of $\sigma_\mathrm{tot}=
0.4$~km~s$^{-1}$). If this is true, then the star formation which
is clearly occurring in the filaments must also happen on a crossing time.

Given filament masses of 5--11~\msun\ per 0.1~pc, there is no requirement
for substantial large scale flows along or onto the filaments in order
to gather sufficient material for further star formation.  Typically,
in regions where cores have already formed, one or two cores are found
within 0.1~pc, and local fragmentation will suffice to explain their
masses. On the other hand, each filament does not hold a vast reservoir
of mass -- a few tens of \msun.  Therefore the filaments are not required
to disperse large fractions of their mass in order to remain consistent
with the observed stellar densities in IC348 or NGC1333.  In other words
the conversion of dense molecular gas into stars may be fairly efficient
once the filament stage is reached, and most of the filament mass may
ultimately go into cores.  This potentially high conversion efficiency
of dense molecular gas into stars has also been noticed for NH$_3$
cores \citep{fuller&myers87}.  Alternatively, stellar clusters may be
built up by a series of filaments that form a few stars inefficiently,
disperse, and are replaced, requiring ongoing filament production in
the same location.

\section{Column density condition for star formation}
\label{sect:column}

We are interested in the conditions under which stars form and it is
clear that the density of the parent molecular cloud is an important
factor.  Density itself is not so easy to measure directly from dust
continuum or C$^{18}$O emission, but what we can consider is how the
numbers of stars formed varies with the line-of-sight column density
(or equivalently visual extinction).   An extinction threshold for star
formation at $A_{\rm v} = 7$--9 was found in both Taurus and Ophiuchus
\citep{johnstone04, onishi98}, which could indicate that a certain amount
of shielding from the interstellar radiation field is required for star
formation to occur.  \citet{onishi98} measured the column density of
C$^{18}$O associated with sources in Taurus, and found that the cold IRAS
sources (assumed to be the younger protostars) and starless H$^{13}$CO$^+$
cores had a average column density of at least $8\times 10^{21}$~cm$^{-2}$
or $A_{\rm v} = 9$.  \citet{johnstone04} compared 850~$\mu$m cores in
$\rho$~Ophiuchus with an optical/IR extinction map and found no submm
cores below an $A_{\rm v}$ of 7.  In both cases the interpretation is
that star formation is inhibited below a certain $A_{\rm v}$.

The C$^{18}$O integrated intensity is a good tracer of total gas column
density over the typical range of column densities in molecular clouds,
but is less reliable at extreme column densities.  At low column densities
the C$^{18}$O abundance falls due to photodissociation.
   %%
   %% in low density gas.
   %%
On the other hand at high densities the line becomes optically thick
and in addition in low temperature, high density regions the C$^{18}$O
depletes onto dust grains (e.g.~\citealp{tafalla04}) and so ceases to
trace the gas.  But between these extremes, a linear C$^{18}$O -- $A_{\rm
v}$ relationship holds above $A_{\rm v} \sim 4$ \citep{frerking82}.

We see in Fig.~\ref{fig:scubaandCO} that the C$^{18}$O is relatively
smooth on scales of a few arcminutes across the cloud, rather than peaking
at the submm cores.  C$^{18}$O integrated intensity at $1'$ resolution is
a poor tracer of submm cores, tracing instead the environment on larger
scales.  This is probably largely due to beam dilution -- the C$^{18}$O
   effective beam area
   is $(1'/14'')^2 =18$~times greater than the 850~\micron\ beam,
    so a core of as much as 0.7~\msun\ could produce
     less than 1~K~km~s$^{-1}$ in $\hbox{C}^{18}\hbox{O}$ integrated
     intensity if it had no extended envelope.   Depletion
   also helps reduce the contribution of core emission to the
   C$^{18}$O.  C$^{18}$O depletion has been measured in prestellar
   cores on scales of 0.1~pc \citep{tafalla04}, with the C$^{18}$O
   abundance substantially reduced in the dense parts of the cores.
   Finally, the C$^{18}$O~1--0 line saturates at about the same column
   density at which the SCUBA 850~\micron\ emission becomes visible,
   which further reduces the contribution from the highest column
   density lines-of-sight.  The result is that the C$^{18}$O traces
   the environment of the cores, and thus, if we look at young enough
   cores, the environment in which cores form.

\subsection{Probability of finding a core}

In this section we investigate how the probability of finding a core
varies with the C$^{18}$O integrated line intensity on arcminute scales.

%%
%% I've fixed the formatting of the following table to match A&A style -- CJC
%%
\begin{table}
\caption{Fraction of cloud by area and number of cores in each
C$^{18}$O integrated intensity range.}
\label{tbl:ncores}
\begin{tabular}{lcccc}
\hline\hline
$I(\,\hbox{C}^{18}\hbox{O})$ & $A_{\rm v}$ &\% area$^1$ &
                             $N$(cores) & \% of cores\\
K~km~s$^{-1}$ & & & &\\
\hline
$<1$ &$<5.2$  &83 &10 &11.4\\
1--2 &5.2--8.8  &13 &14 &15.9\\
2--3 &8.8--12.4  &3  &32 &36.4\\
3--4 &12.4--15.9  &0.5 &17 &19.3\\
4--5 &15.9--19.5  &0.2 &13 &15.9\\
$>5$ &$>19.5$  &0.04 &1 &1.1\\
\hline
\end{tabular}
\begin{list}{}{}
\item[$^1$]Fraction of the area of the cloud mapped in C$^{18}$O
\end{list}
\end{table}

\begin{figure}[t]
\includegraphics[scale=0.35,angle=-90]{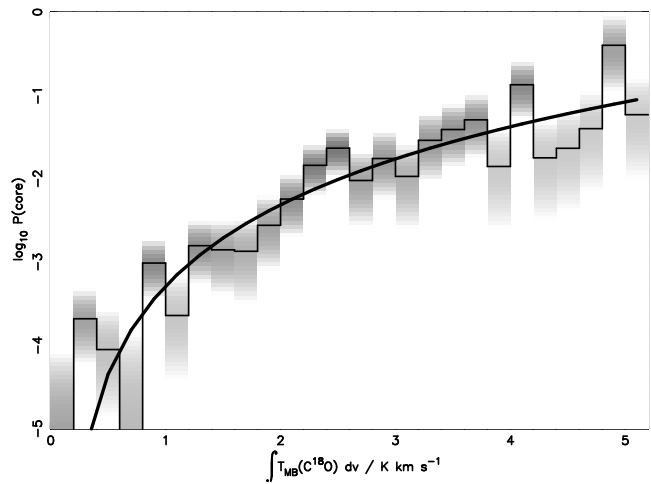}
\caption{Probability of finding a submm core in a $22''$ square pixel with
   a given $1'$ beam-averaged C$^{18}$O integrated intensity: observed (thin
   black line)and probability distribution of $\log_{10}(\prob(\hbox{core}))
   = N_\mathrm {cores}(I_i)/N_\mathrm{pixels}(I_i)$ for each integrated
   intensity bin $I_i$ (greyscale), and best-fit power law (thick black
   line).  To convert the probabilities to another pixel size use $P' =
   P(\hbox{core})\times{\hbox{(pixel area)}\over22^2}$.}
\label{fig:pcore}
\end{figure}

Table~\ref{tbl:ncores} compares the area of the cloud in each C$^{18}$O
integrated intensity range with the number of submm cores in that region.
Although there are clearly cores associated with areas of the cloud
with a wide range of cloud column densities, cores are much more likely
to be found in regions of high column density.  In Fig.~\ref{fig:pcore}
we demonstrate this graphically by plotting the probability of finding a
submm core, $\prob$(core), as a function of integrated C$^{18}$O
intensity, $I(\,\hbox{C}^{18}\hbox{O})$ (K~km~s$^{-1}$).  The probability
of finding a core is derived from two observed distributions.  The first
is the C$^{18}$O integrated intensity towards the peak of each submm
core, determined from the C$^{18}$O map (Fig.~\ref{fig:scubaandCO}b).
We bin the results by C$^{18}$O integrated intensity to generate $N_{\rm
cores} (I_i)$, the number of submm cores in each integrated intensity
bin.  The second is a count of the number of pixels in the map in each
integrated intensity bin, $N_{\rm pixels}(I_i)$.  This gives a measure
of the (probability) distribution of integrated intensities $\prob(I)$
in the surveyed cloud,  or in other words how much of the cloud area lies
in a particular integrated intensity range.  The probability of finding
a submm core at a position with C$^{18}$O integrated intensity $I$ is
then given by $\prob(\hbox{core}) = N_{\rm cores}(I)/N_{\rm pixels}(I)$.

As both distributions $N_{\rm cores}(I_i)$ and $N_{\rm pixels}(I_i)$
at some point contain very small numbers of counts we have to be very
careful with the error budget in calculating this quotient.  We take
the Bayesian approach detailed in Appendix~\ref{appendix} to calculate
$\prob$(core).

Fig.~\ref{fig:pcore} shows that $\prob(\hbox{core})$ is a steeply
increasing function of integrated intensity $I$.  In other words,
the probability of finding a core increases rapidly with C$^{18}$O
column density.  Fitting (for simplicity, and for want of a theoretical
prediction) a power law $kI^\alpha$ to the distribution gives a best fit
result of $\alpha = 3.00_{-0.21}^{+0.18}$, $log_{10}\,k = -3.15\pm0.08$
(uncertainties are 68.3\% confidence limits).  $k$ is the scaling factor
which gives the probability of finding a core in a $22''$ square pixel.
A power law must break down as a description of $\prob(\hbox{core})$ at
high $I(\,\hbox{C}^{18}\hbox{O})$ as the probability of finding a core
in any one pixel approaches 1 but this limit is not approached with our
pixel size.

 The exact shape of the distribution is uncertain at the low end
because of incomplete map coverage and lack of sensitivity.  Both JCMT
and FCRAO maps have limited coverage of low column density regions,
and as the area mapped in C$^{18}$O is larger, we may have missed
SCUBA cores. So we may underestimate (by only a few cores, we believe)
the probability of finding a core below $I(\,\hbox{C}^{18}\hbox{O}) =
1.0 \hbox{ K km s}^{-1}$.
Fig.~\ref{fig:scubaandCO} also shows that the C$^{18}$O observations
were not biassed towards regions with SCUBA clumps, so we do not
overestimate the number of cores at low intensity.
   The measurement uncertainties also become significant below
   $I\,(\hbox{C}^{18}\hbox{O}) = 1\hbox{ K km s}^{-1}$ ($A_{\rm v} =
   5.2$) -- the C$^{18}$O RMS is 0.3~K~km~s$^{-1}$.   Therefore, the source count in the lower 5 integrated intensity
   bins (Fig.~\ref{fig:pcore}) is also uncertain.  We made a
   fit excluding the lowest column density channels
   ($I\,(\mathrm{C}^{18}\mathrm{O}) < 1.0\hbox{ K km s}^{-1}$), which
   steepens the power law to $\alpha = 3.45\pm0.27$, a change which is
   significant only at the 95\% level.

What does $\prob(\hbox{core}) \propto I\,^{3.0}$ mean?  We can rule
out an absolute threshold in $N(\,\hbox{C}^{18}\hbox{O}$) (and by
implication $A_{\rm v}$) for submm cores (though not necessarily for
star formation).  There are clearly real submm cores at low C$^{18}$O
integrated intensities of 4~K~km~s$^{-1}$ or visual extinctions as low
as $A_{\rm v}=3$ ($N$(H$_2) = 2.7\times 10^{21}$~cm$^{-2}$).  At these
column densities the uncertainty on the C$^{18}$O data is high but the
low extinctions are independently confirmed by $^{13}$CO data from the
same observing run.  There are submm cores at very low C$^{18}$O column
density despite the fact that the peak column densities in the cores must
be high to be detected with SCUBA, and that the observations are less
sensitive to the larger, more diffuse cores which form in low pressure
regions (Sect.~\ref{sect:observations}).

The 10\% of cores with $I < 1.0$~K~km~s$^{-1}$ include the L1455 cluster
plus IRAS~03235+3004, 03422+3256, and 03262+3123.  The question remains
whether stars can form at these low column densities or if these
low column density cores are all more evolved (Class~I) protostars,
or whether the lower angular resolution of the line data is biasing
the column density estimate to a low value.  Certainly the majority
of the $I < 1$~K~km~s$^{-1}$ ($A_{\rm v} < 5.2$) sources have IRAS
identifications and are therefore likely to be classified as Class~I.
Potentially less evolved cores are L1455~smm1
   ($\alpha(2000)=03^{\rm h} 27^{\rm m} 42\fs9$,
   $\delta(2000)=+30^\circ 12' 28'')$, which has a CO outflow, is
   definitely protostellar, and has low IRAS fluxes, L1455~smm2
($\alpha(2000)=03^{\rm h} 27^{\rm m} 46\fs6$,
$\delta(2000)=+30^\circ 12' 05''$; probably starless),
and three weak cores to the west of B1 which may be noise artefacts.
\citet{onishi98} found that cold cores in Taurus were only found at a
high C$^{18}$O threshold of $8 \times 10^{21}$~cm$^{-2}$, equivalent
to $I(\,\hbox{C}^{18}\hbox{O}) \sim 2.0$~K~km~s$^{-1}$ or $A_{\rm v} =
9$, whereas warm cores (more evolved IRAS sources) could appear at lower
C$^{18}$O column density.  This appears to roughly hold in Perseus as well
as Taurus, but we need to investigate this further once we can better
classify the Perseus sources.  All the submm cores in Ophiuchus lie at
$A_{\rm v} \geq 7$ \citep{johnstone04}.   As the cores in Ophiuchus were
identified in the same way as in this study of Perseus, that is, from
SCUBA 850~$\mu$m observations, there must either be a real difference
between star formation in the two clouds, or a significant difference in
the way that optical/IR extinction and C$^{18}$O trace column density in
the range $A_{\rm v} = 2$--7.  $\rho$~Oph is lacking in starless cores and
Class 0 protostars compared to other dark clouds and there may have been
a burst of star formation $10^5$~years ago in this region that has now
tailed off \citep{visser02}; in Perseus, however, we have clear evidence
for continuing star formation, so there are differences between the ages
of the populations in the two clouds which might underly the differing
column density thresholds, though the mechanism is unclear.

The $I^{3.0}$ power law rules out  simple models in which the
number of cores is proportional to the mass along the line-of-sight
($\prob(\hbox{core}) \propto I$).  If density is the determining quantity
for star formation, central densities must rise steeply with increasing
column density.  Theoretical models which reproduce the observed C$^{18}$O
integrated intensity -- core probability relationship are required to
find the actual physics behind the $I\,^{3.0}$ power law (if indeed a
power law is the right functional form).

This steep rise in $\prob$(core) with column density is consistent
with the overall trend seen in star forming regions for high column
density regions to be associated with formation of larger numbers
of stars.  Isolated low mass star formation in Taurus occurs at
column densities of $\simeq 8\hbox{--}15\times 10^{21}$~cm$^{-2}$
\citep{onishi98}.  Column densities for clustered star formation in
Perseus reach $2\times10^{22}$~cm$^{-2}$.  In contrast, column densities
in massive star forming regions typically exceed $10^{23}$~cm$^{-2}$
\citep{hatchell98linesurvey}.

We have already mentioned C$^{18}$O depletion and optical depth, both
of which act to reduce the observed $I(\,\hbox{C}^{18}\hbox{O})$ for a
given H$_2$ column density.   If plotted against H$_2$ column density,
the $\prob$(core) distribution would flatten at high column
densities.   A rough correction for the optical depth suggests that
the power law index could reduce from $3.0$ to $\sim 2.0$ in the worst
case.  This is still significantly steeper than linear.

Given a C$^{18}$O map and the above P(core) vs.
$I(\,\hbox{C}^{18}\hbox{O})$ relationship, we can now predict --
at least statistically -- how many submm cores will be found and how
they will be distributed.  It will be interesting to see how well this
relationship holds in regions other than Perseus.  There are obviously
more factors involved in core formation than C$^{18}$O column density,
and a better model of where stars form will need to take into account
filaments, clustering and other factors.

\section{Conclusions}
\label{sect:conclusions}

We have mapped the submm dust emission from the Perseus molecular
cloud with SCUBA at JCMT with 5000~AU resolution, revealing 91 embedded
starless and protostellar cores and the filamentary structures which
lead to star formation.   The total mass of the Perseus cloud derived
from extinction measurements is 17,000~\msun\ \citep{bachiller86av}.
Of this, 6000~\msun\ is traced by C$^{18}$O and 2600~\msun\ by dust in
cores and clusters.  We conclude:

\begin{enumerate}

\item Only a small fraction, less than 20\%, of the mass of the molecular
   cloud is involved in star formation at this time.
\item By number, 80\% of the star formation is occuring in clusters,
   defined by a stellar density of more than 1~\msun~pc$^{-3}$.  Some
   current star formation activity may be associated with the larger
   stellar clusters IC348 and NGC1333, but about half the star formation
   is taking place in small clusters
containing less than 35 members.
\item Filamentary structure is evident both in the molecular gas on
   20~pc scales and on small scales in the dust emission.  High column
   density dust filaments with masses per unit length of 5--11~\msun\
   per 0.1~pc are associated with the existing clusters.  This range
   of masses is in agreement with models of filaments with either
   unmagnetised filaments or a slightly dominant toroidal field
   \citep{fiege&pudritz00}.
\item The probability of finding a submm core is a steeply rising function
   of column density (as measured by C$^{18}$O integrated intensity):
   $\prob(\hbox{core}) \propto I\,^{3.0}$. There is no C$^{18}$O column
   density cutoff (and by implication $A_{\rm v}$) below which there
   are no cores.

\end{enumerate}

These
   results are specific to Perseus, and
it will require similar studies of a much larger sample of such clouds
before we can fully understand which of these features are typical of
star forming clouds and which are, for some reason, unique to Perseus.

\acknowledgements

We would like to thank Jane Buckle, Tak Fujiyoshi and others who spent
long nights at the telescope collecting data as part of this project,
and the referee Doug Johnstone for his careful reading and constructive
suggestions.  The James Clerk Maxwell Telescope is operated by the
Joint Astronomy Centre on behalf of the Particle Physics and Astronomy
Research Council of the United Kingdom, the Netherlands Organisation
for Scientific Research, and the National Research Council of Canada.
The FCRAO is supported by the NSF via AST-0100793.  JH acknowledges
support from DFG SFB 494 and the PPARC Advanced Fellowship programme.
The National Radio Astronomy Observatory is a facility of the National
Science Foundation operated under cooperative agreement by Associated
Universities, Inc.

\appendix

\section{Details of power law fitting procedure}
\label{appendix}

In this Appendix we give details of the analysis of the core probability
vs. C$^{18}$O column density.  We first assume that the counts in each
integrated intensity bin $I_i$,  $x_i = N_\mathrm{cores}(I_i)$ and $y_i =
N_\mathrm{pixels}(I_i)$, obey Poisson statistics, i.e., are related to
the true values $\mu_x$ and $\mu_y$ by:
$$\prob(x\,|\,\mu_x) = {\mu_x^x e^{-\mu_x}\over
x!},\qquad\prob(y\,|\,\mu_y) = {\mu_y^y e^{-\mu_y}\over y!}$$

By Bayes' theorem we can reverse this.  We take a a flat prior with a
positivity constraint on $\mu_x$ and $\mu_y$, $\prob(\mu_x) = 1\,(\mu_x
\geq 0) \hbox{ or } 0\, (\mu_x < 0)$ and similarly for $\mu_y$ (see, e.g.,
\citealp{sivia96} for an introduction to Bayesian data analysis).  Then
\begin{eqnarray}
\label{eqn:x}
\prob(\mu_x\,|\,x) &= \prob(x\,|\,\mu_x)\times\prob(\mu_x) = {\mu_x^x
e^{-\mu_x}\over x!}\\
\prob(\mu_y\,|\,y) &= \prob(y\,|\,\mu_y)\times\prob(\mu_y) = {\mu_y^y
e^{-\mu_y}\over y!}\label{eqn:muy}
\end{eqnarray}
and similarly for $\mu_y$.  

We calculate the data-derived probability of finding a core at
integrated intensity $I_i$, $C_{{\rm obs},i} = x_i/y_i$ directly.
However, this is only an estimate of the true value of $C$ because of
the uncertainties in $x_i$ and $y_i$.  We next calculate the PDF for
the true value of finding a core $\prob(C_{\rm true}\,|\,x,y)$ given
$x_i$ and $y_i$.  To obtain the probability distribution of the true
quotient $C_{\rm true}$ given the observed $x$ and $y$ we rewrite the
$x$ distribution (Equation~\ref{eqn:x}) in terms of $\mu_y$ and
$C_{\rm true}$:
\begin{equation}\prob(\mu_y\,|\,x,C_{\rm true}) = {(C_{\rm true})^x e^{-C_{\rm
true}\mu_y}\over x!}.\label{eqn:muy2}\end{equation}

Now with two probability distributions on $\mu_y$ (Equations~\ref{eqn:muy} and~\ref{eqn:muy2}) we can multiply and
integrate out $\mu_y$:
\begin{eqnarray}
   \prob(C_{\rm true}\,|\,x,y) &\propto& \int_{0}^{\infty} {{\mu_y^y
   e^{-\mu_y}\over y!} {(C_{\rm true})^x e^{-C_{\rm true}\mu_y}\over x!}
   d\mu_y }\nonumber\\
   &\propto& {(x+y)^{(x+y)}\over (1+C_{\rm true})^{(x+y+1)}}{C_{\rm
   true}^x\over x! y!}.\nonumber
\end{eqnarray}

\begin{figure}[t]
\includegraphics[scale=0.35,angle=-90]{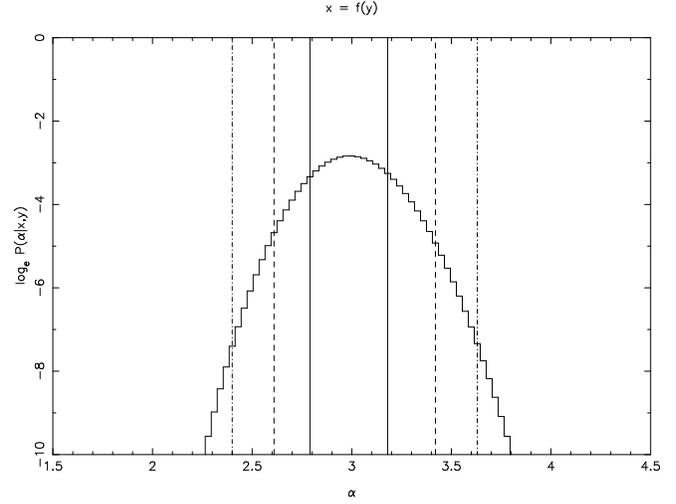}
\caption{Probability density function for the power law index $\alpha$,
   $\log(\prob(\alpha|x,y))$.  Verticals are 68.3\%,95.4\% and 99.7\%
   confidence limits.}
\label{fig:palpha}
\end{figure}

Finally to find the most probable power law fit, for each $I_i$ we
calculate $\prob(C_{\rm true},i = kI_i^\alpha\,\,|\,\,x_i,y_i)$
from the previous formula and $$\prob(k,\alpha\,\,|\,\,x,y)
=\prod\prob(k,\alpha\,\,|\,\,x_i,y_i).$$ The PDFs for $C_{\rm true}$
are plotted as greyscales and contours for each integrated intensity
bin on Fig.~\ref{fig:pcore}.  We marginalise (integrate over) $k$
and $\alpha$ in turn to estimate the uncertainties in each parameter
individually as given in Sect.~\ref{sect:column}.  The PDF for $\alpha$
is shown in Fig.~\ref{fig:palpha}.

\bibliographystyle{aa}
\bibliography{perseus}
%\bibliography{1836}

\onecolumn
\setlength\LTleft{0pt}
\setlength\LTright{0pt}

\newpage
\begin{landscape}
\include{SSA-clusters-sorted-tidy}
\end{landscape}
\twocolumn

\end{document}

%% file: SSA-clusters-sorted-tidy.tex
%%
%% I've fixed the formatting of the following table to match A&A style -- CJC
%%
\begin{longtable}{cl@{:}l@{:}ll@{:}l@{:}llllcc} 
\caption{SCUBA 850~$\mu$m core positions, names and references, peak
850~$\mu$m fluxes and C$^{18}$O integrated intensity at the submm peak
position.\label{tbl:sourcelist}}\\
\hline\hline
No. & \multicolumn{3}{c}{RA$_{2000}$} & \multicolumn{3}{c}{Dec$_{2000}$} &
Cluster$^{(1)}$ & Name & Reference$^{(2)}$ & $F_{850}^{(3)}$ &
$I(\hbox{C}_{18}\hbox{O})^{(4)}$\\
 &\multicolumn{3}{c}{hh:mm:ss} & \multicolumn{3}{c}{dd:mm:ss}& & & & mJy/bm &
 K~km~s$^{-1}$ \\
\hline
\endfirsthead
\caption{continued.}\\
\hline\hline
No. & \multicolumn{3}{c}{RA$_{2000}$} &
\multicolumn{3}{c}{Dec$_{2000}$} &
Cluster$^{(1)}$ & Name & Reference$^{(2)}$ & $F_{850}^{(3)}$ & 
$I(\hbox{C}_{18}\hbox{O})^{(4)}$\\
 &\multicolumn{3}{c}{hh:mm:ss} & \multicolumn{3}{c}{dd:mm:ss}& & & &
mJy/bm &    K~km~s$^{-1}$ \\
\hline
\endhead
\hline
\endfoot
1&03&33&17.9&31&09&33&B1&B1-c&\citet{matthews02}&1638&4.0\\ 
2&03&33&21.4&31&07&31&B1&B1-bS&\citet{hirano99,matthews02}&1554&4.8\\ 
3&03&33&21.1&31&07&40&B1&B1-bN&\citet{hirano99,matthews02}&1200&4.8\\ 
4&03&33&16.3&31&06&54&B1&B1-d&\citet{matthews02}&514&4.1\\ 
5&03&33&01.9&31&04&23&B1&None&&218&3.3\\ 
6&03&33&05.6&31&05&06&B1&None&&207&2.8\\ 
7&03&33&16.5&31&07&51&B1&B1-a&\citet{hirano99,matthews02}&202&4.5\\ 
8&03&33&04.0&31&04&56&B1&None&&202&2.9\\ 
9&03&33&03.0&31&04&38&B1&None&&196&3.4\\ 
10&03&33&27.3&31&07&10&B1&None&&179&4.2\\ 
11&03&33&00.3&31&04&17&B1&None&&160&3.4\\ 
12&03&43&56.5&32&00&50&HH211&HH211&\citet{mccaughrean04,chandlerricher00}&1494&2.5\\ 
13&03&43&56.9&32&03&05&HH211&IC348 MMS&\citet{eisloeffel03}&1032&3.1\\ 
14&03&44&43.9&32&01&32&HH211&None&&623&2.0\\ 
15&03&30&15.5&30&23&43&HH211&No SMM/MM source&&358&2.2\\ 
16&03&44&01.0&32&01&55&HH211&None&&296&2.4\\ 
17&03&43&57.9&32&04&01&HH211&None&&247&3.0\\ 
18&03&44&03.0&32&02&24&HH211&No SMM/MM source&&237&2.5\\ 
19&03&44&36.8&31&58&49&HH211&None&&234&1.8\\ 
20&03&44&05.5&32&01&57&HH211&No SMM/MM source&&209&2.4\\ 
21&03&44&02.3&32&02&48&HH211&None&&207&2.9\\ 
22&03&44&06.2&32&02&12&HH211&None&&204&2.4\\ 
23&03&43&37.8&32&03&06&HH211&None&&170&1.5\\ 
24&03&43&42.5&32&03&23&HH211&None&&168&2.3\\ 
25&03&44&48.5&32&00&31&HH211&None&&153&2.0\\ 
26&03&43&44.4&32&02&56&HH211&None&&153&2.5\\ 
27&03&25&35.9&30&45&30&L1448&L1448NW     &\citet{looney00,barsony98}&2300&2.9\\ 
28&03&25&36.4&30&45&15&L1448&L1448N B  &\citet{looney00,barsony98}&2243&2.7\\ 
29&03&25&38.8&30&44&04&L1448&L1448 C&\citet{bachiller91,barsony98}&1737&2.3\\ 
30&03&25&22.4&30&45&11&L1448&L1448-IRS2  &\citet{olinger99}&1316&2.2\\ 
31&03&25&25.9&30&45&03&L1448&None&&429&2.4\\ 
32&03&25&49.0&30&42&25&L1448&None&&364&2.0\\ 
33&03&25&50.6&30&42&01&L1448&None&&306&2.1\\ 
34&03&25&30.8&30&45&07&L1448&None&&231&2.6\\ 
35&03&27&39.1&30&13&01&L1455&RNO 15 FIR&\citet{rengel02}&556&1.4\\ 
36&03&27&42.9&30&12&28&L1455&None&&402&0.9\\ 
37&03&27&48.4&30&12&09&L1455&No SMM/MM source&&324&0.8\\ 
38&03&27&46.6&30&12&05&L1455&No SMM/MM source&&272&0.9\\ 
39&03&27&38.1&30&13&57&L1455&No SMM/MM source&&238&1.4\\ 
40&03&27&39.9&30&12&10&L1455&None&&232&1.4\\ 
41&03&29&10.4&31&13&30&NGC1333&IRAS 4A&\citet{sandell91}&7000&3.7\\ 
42&03&29&12.0&31&13&10&NGC1333&IRAS 4B; SK-3&\citet{sandell91,sandellknee01}&3025&3.4\\ 
43&03&29&03.2&31&15&59&NGC1333&HH7-11 MMS 1;SK-12/13&\citet{chini97}&2831&4.0\\ 
44&03&28&55.3&31&14&36&NGC1333&IRAS 2A; SK-8&\citet{sandell94,sandellknee01}&2355&3.8\\ 
45&03&29&01.4&31&20&29&NGC1333&SK-24&\citet{sandellknee01}&1119&3.7\\ 
46&03&29&10.9&31&18&27&NGC1333&SK-20&\citet{sandellknee01}&1061&3.6\\ 
47&03&28&59.7&31&21&34&NGC1333&SK-31&\citet{sandellknee01}&770&4.9\\ 
48&03&29&13.6&31&13&55&NGC1333&IRAS 4C; SK-5&\citet{looney00,sandellknee01}&667&3.3\\ 
49&03&28&36.7&31&13&30&NGC1333&SK-6&\citet{sandellknee01}&621&2.0\\ 
50&03&29&06.5&31&15&39&NGC1333&HH 7-11 MMS 4;SK-15&\citet{chini01,sandellknee01}&619&3.2\\ 
51&03&29&08.8&31&15&18&NGC1333&SK-16&\citet{sandellknee01}&574&2.8\\ 
52&03&29&03.7&31&14&53&NGC1333&HH7-11 MMS 6; SK-14&\citet{chini01,sandellknee01}&517&3.8\\ 
53&03&29&04.5&31&20&59&NGC1333&SK-26&\citet{sandellknee01}&513&4.8\\ 
54&03&29&10.7&31&21&45&NGC1333&SK-28&\citet{sandellknee01}&420&3.6\\ 
55&03&28&40.4&31&17&51&NGC1333&None&&410&1.9\\ 
56&03&29&07.7&31&21&57&NGC1333&SK-29&\citet{sandellknee01}&394&4.3\\ 
57&03&29&18.2&31&25&11&NGC1333&SK-33&\citet{sandellknee01}&376&4.1\\ 
58&03&29&24.0&31&33&21&NGC1333&None&&371&0.9\\ 
59&03&29&16.5&31&12&35&NGC1333&None&&371&3.2\\ 
60&03&28&39.3&31&18&27&NGC1333&None&&354&2.9\\ 
61&03&29&17.3&31&27&50&NGC1333&None&&349&2.3\\ 
62&03&29&07.1&31&17&24&NGC1333&SK-18&\citet{sandellknee01}&318&4.0\\ 
63&03&29&18.8&31&23&17&NGC1333&SK-32&\citet{sandellknee01}&308&3.8\\ 
64&03&29&25.5&31&28&18&NGC1333&None&&292&1.4\\ 
65&03&29&00.4&31&12&01&NGC1333&SK-1&\citet{sandellknee01}&288&2.3\\ 
66&03&29&05.3&31&22&11&NGC1333&SK-30&\citet{sandellknee01}&228&5.1\\ 
67&03&29&19.7&31&23&56&NGC1333&None&&223&4.7\\ 
68&03&28&56.2&31&19&13&NGC1333&No SMM/MM source&&201&3.9\\ 
69&03&28&34.4&31&06&59&NGC1333&None&&201&1.3\\ 
70&03&29&15.3&31&20&31&NGC1333&SK-22&\citet{sandellknee01}&192&4.2\\ 
71&03&28&38.7&31&05&57&NGC1333&None&&177&1.9\\ 
72&03&29&19.1&31&11&38&NGC1333&No SMM/MM source&&167&2.6\\ 
73&03&28&38.8&31&19&15&NGC1333&None&&162&2.4\\ 
74&03&28&32.5&31&11&08&NGC1333&None&&156&1.9\\ 
75&03&28&42.6&31&06&10&NGC1333&None&&153&2.2\\ 
76&03&32&17.8&30&49&46&None&No SMM/MM source&&1916&1.7\\ 
77&03&31&21.0&30&45&28&None&I03282+3035 &\citet{bachiller94,motteandre01}&892&1.1\\ 
78&03&47&41.6&32&51&44&None$^{5}$&B5 IRS1&\citet{motteandre01}&424&?\\ 
79&03&47&39.1&32&52&18&None$^{5}$&None&&335&?\\ 
80&03&26&37.6&30&15&24&None&No SMM/MM source&&302&0.2\\ 
81&03&43&50.8&32&3&24&None&No SMM/MM source&&253&0.9\\ 
82&03&33&13.1&31&19&51&None&None&&219&2.4\\ 
83&03&32&48.9&31&09&40&None&None&&188&0.3\\ 
84&03&32&21.9&31&04&56&None$^{6}$&None&&184&0.2\\ 
85&03&28&32.5&31&00&53&None&03254+3050&\citet{dent98}&177&1.5\\ 
86&03&26&30.9&30&32&28&None&None&&176&1.5\\ 
87&03&32&21.5&31&05&08&None$^{6}$&None&&166&0.4\\ 
88&03&31&31.6&30&43&32&None&None&&165&2.2\\ 
89&03&32&25.9&30&59&05&None&None&&164&2.2\\ 
90&03&45&16.5&32&04&47&None&None&&163&0.8\\ 
91&03&29&23.3&31&36&08&None&None&&154&?\\ 
\hline
\multicolumn{12}{p{18.0cm}}{$^{(1)}$ If the core satisfies the
cluster membership criterion (Sect.~\ref{sect:clusters}) then the
cluster name is given here.}\\
\multicolumn{12}{p{18.0cm}}{$^{(2)}$ References are to the earliest
previous detection of the submillimetre source.   Full
cross-identifications with sources at other wavelengths are
non-trivial due to differing beam sizes and are left for a future
analysis, but to indicate an obvious likely counterpart (e.g., IRAS),
`No mm/smm source' is marked in the `Name' column instead of `None'.}\\
\multicolumn{12}{p{18.0cm}}{$^{(3)}$ SCUBA 850~$\mu$m peak flux ($14''$ beam)}\\
\multicolumn{12}{p{18.0cm}}{$^{(4)}$ C$^{18}$O integrated intensity
($1'$ beam averaged) at the 850~$\mu$m peak.  A question mark in
this column indicates a SCUBA peak which lies outside the C$^{18}$O map.}\\
\multicolumn{12}{p{18.0cm}}{$^{(5)}$ B5 pair }\\
\multicolumn{12}{p{18.0cm}}{$^{(6)}$ B1 west pair}\\
\end{longtable}